\documentclass[12pt]{article}

\usepackage{ucs}
\usepackage[utf8x]{inputenc}
\usepackage[english]{babel} 
\usepackage{amsmath}
\usepackage{amsfonts}
\usepackage{graphicx}
\usepackage{epstopdf}

\title{Phonon transmission across an interface between two crystals}
\author{A.\,P.\,Meilakhs}
\date{\today}

\begin{document}
  \maketitle

\section*{Abstract}
     The new model of phonon transmission across the interface between two crystals is proposed featured by taking into account the mismatch of crystal lattices. It has been found that the mismatch of lattices results in phonon scattering at the interface even in the absence of defects. As it has been shown, at the normal incidence, longitudinally polarized phonons have much larger transmission coefficient than that of transversely polarized phonons, excluding the special resonance cases. For the quasi one-dimensional case the exact solution has been obtained.

\section{Introduction}

When the heat flows through the boundary between two crystals, temperature at the interface experiences a sharp jump. The proportionality coefficient between the heat flux and the temperature jump is known as the thermal boundary resistance or the Kapitza resistance. The theory of a Kapitza resistance attracts recently serious attention of researchers due to its significance for practical use \cite{ShK, ZhW, Ter, Wang}. The study of thermal transmission across the crystal interface, in its own turn, attracts the attention to the problem of phonon transmission across the crystal interface since, in the majority of cases, the phonons are responsible for the energy transfer across the interface.

     Various approximations are used in the theory of thermal boundary resistance to describe the dynamics of the crystal lattice in the boundary region near the interface between two crystals. Two basic approaches are so called Acoustic Mismatch Model (AMM) and Diffusive Mismatch Model (DMM). In the first approach \cite{Kh, Can} it is suggested that the dynamics of a lattice and, particularly, the interfacial coefficients of the phonon transmission and reflection could be calculated using the elasticity theory. However, firstly, this approximation is suitable only for a calculation of the Kapitza resistance at low temperatures since in this case only low frequency crystal oscillations are excited, which are properly well described by the elasticity theory. Secondly, this approximation doesn’t take into account the phonon scattering at the crystal interface. The values of the Kapitza resistance calculated within AMM occurred significantly higher than the experimentally determined values \cite{St}. 

     Second approach \cite{Swar} suggests, vice versa, that the interfacial scattering is very strong, and phonons incident at the interface “forget” their initial direction and uniformly scatter in all directions. The Kapitza resistance calculated within this approach is higher than the value calculated within the AMM, but still doesn’t give the good agreement with the experimental data \cite{St}. Besides, the authors of \cite{St} have studied experimentally the dependence of the Kapitza resistance on surface roughness. It turned out that the better the surface is polished and, hence, the lower the scattering at it, the higher the thermal transport through the surface. But this conclusion contradicts the results of the DMM calculations.  Thus, the DMM approximation could be regarded as refuted.          

Thus, it is necessary to develop more accurate model of the phonon transmission across the interface, which takes into account the atomic structure of actual crystals. The simplest model of this kind is the one-dimensional chain with the interface. Such a model has been detailed in papers \cite{Zh, Me}. Though, since an actual crystal is the three-dimensional structure, the one-dimensional chain permits to elucidate only some qualitative properties of the lattice dynamics connected with the atomic structure. The paper \cite{Young} describes the three-dimensional model of the lattice dynamics near the interface. However, this model does not take into consideration the lattice mismatch of two crystals.

    More complicated models are studied by computer simulation. Thus, in paper \cite{Hu} it was found that both the AMM and DMM do not give the proper description of the phonon transmission across the interface. It has been also found that the longitudinal phonons have the significantly higher interfacial transmittance coefficient than the transverse phonons. 
The phonon transmission across the interface is studied with a Green’s function method in \cite{Chen}. It is also suggested that atoms located at the interface are randomly displaced from the position that they would occupy if they were in volume of the crystal. 

The role of anharmonicity of oscillations in the phonon transmission across the interface is considered in the paper \cite{Anh}. In the most recent study \cite{Nuo}, dealing with the numerical simulation of the lattice dynamics at the crystal interface, it has been found that atoms at the interface oscillate with frequency exceeding the maximum possible frequency in a given crystal.

    The presented paper proposes the analytical description of dynamics of the crystal lattice near the interface, which takes into account the mutual mismatch of the lattices. The main assumption is that displacements of atoms at the interface are not random, but determined by forces affecting them from the side of atoms of the adjacent crystal. Generally the exact solution in such a model is not possible. At the same time, the essential qualitative effects related to the lattice mismatch can be found. In particular, it is possible to explain why the transverse phonons have much lower transmission coefficients compared to the longitudinal ones. It turns out that the lattice mismatch, even in the absence of defects, results in phonon scattering at the interface. This scattering is not random, but has a certain structure.

\section{The model}

We consider the interface between two crystals having the structure of a simple cubic lattice and contacting by crystal surfaces \\ $(1, 0, 0)$ (Fig. 1). The axis $x$ is normal to the interface. In each of the crystals we take into account the interaction with the atoms of the first and second coordination groups. Such a model of a three-dimensional crystal lattice is the most simple and well-studied \cite{Ans}.   
               
     The lattice constant for the crystal on the left ($x$ is less than zero) and on the right ($x$ is larger than zero) is $a^L , a^R$, consequently. The constant of the quasi elastic coupling with the atoms of the first and second coordination groups for the atoms of the left crystal and for the atoms of the right crystal is $\beta^L_1, \beta^L_2$, consequently. We neglect the changes of the quasi elastic coupling constants and lattice constants near the interface. It is suggested that the crystals are infinite, the left crystal occupies a half space $x<0$ while the right crystal a half space $x>0$. It is also assumed that interaction between the atoms is due to short-range forces, and that with the atoms of the opposite crystal interact only the atoms lying at the interface.

     We introduce the following numbering: $n_x$ is the numbering of atoms along the axis $x$. For the left crystal the numbering goes from minus infinity to zero. Atoms in the interface plane are numbered as zeroes. For the right crystal the numbering goes from zero to plus infinity, number zero is for atoms in the interface plane. $n_y, n_z$  is the numbering of atoms in the interface plane, it goes from the minus to plus infinity. The bold marking denotes a set of indexes $\mathbf{n} = (n_x, n_y, n_z)$.

\begin{figure}
\centering
\includegraphics[width=0.6\textwidth]{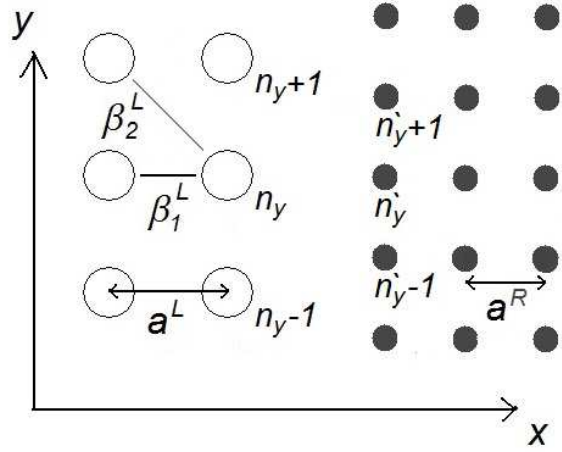}
\caption{ Shown are the interface between two crystals (viewed from the side) and the atomic bonds considered in the model. The lattice constants are denoted as $a_1 , a_2$; the interfacial atoms are numbered.}
\end{figure}

Atoms located near the interface are influenced by an external potential produced by atoms from the opposite side of the interface. Due to this influence, the crystal is deformed, and the equilibrium position of an atom changes relative to its positions in an ideal crystal. Following the theory of an ideal crystal lattice without interface, we could expand the potential energy in the Tailor series in displacements of atoms from equilibrium positions and take into account only the minor terms. The Hamiltonian thus obtained would not possess the symmetry, and the analytical solution would be impossible. Instead, we perform the expansion in displacements of atoms from the positions which the nearest-to-interface atoms would occupy if they would not interact with atoms of an adjacent crystal. At this approach, the terms of the Tailor series containing the first derivative are not equal to zero as it would be the case when the expansion is produced in the neighborhood of the true minimum of the potential energy. The derivatives of higher than third order, i.e. anharmonic terms, are not taken into consideration. The following expressions are valid for atoms on both sides therefore the index denoted the side is omitted.

 Thus we obtain:
 \begin{equation}
U(.., \vec r_\mathbf{n}, ...) = \sum_{\mathbf{n}, \alpha} \frac{\partial U}{\partial r_{\mathbf{n}, \alpha}} r_{\mathbf{n}, \alpha} + \frac{1}{2} \sum_{\mathbf{n}, \alpha} \sum_{\mathbf{l}, \beta} \frac{\partial^2 U}{\partial r_{\mathbf{n}, \alpha} \partial r_{\mathbf{l}, \beta}} r_{\mathbf{n}, \alpha} r_{\mathbf{l}, \beta},
 \end{equation}
where $r_{\mathbf{n}, \alpha}$-- is the displacement of $\mathbf{n}$-th atom along the axis $\alpha$, $\alpha = x,y,z$; $U(.., \vec r_\mathbf{n}, ...)$  is the potential energy as a function of displacement.

In the equilibrium position the condition $\forall \mathbf{n}, \frac{\partial U}{\partial r_{\mathbf{n}, \alpha}} = 0$ is fulfilled. We substitute the introduced expression for the potential energy into Eq. (1), and obtain 
 \begin{equation}
\frac{\partial U}{\partial r_{\mathbf{n}, \alpha}} + \sum_{\mathbf{l}, \beta} \frac{\partial^2 U}{\partial r_{\mathbf{n}, \alpha} \partial r_{\mathbf{l}, \beta}}  r^0_{\mathbf{l}, \beta}=0.
\end{equation}
This system of $3N$ equations ($N$, number of atoms) determines the displacements $ r^0_{\mathbf{l}, \beta}$ at zero temperature in the absence of atomic oscillations, or 0-displacements, that are displacements of atoms from the position about which the potential energy is expanded into the Tailor series, to the equilibrium position.

Now, let temperature is not equal to zero and atoms oscillate about the equilibrium position. Then the total displacement of an atom consists of the 0-displacement and its displacement due to thermal oscillations: $r_{\mathbf{n}, \alpha} = r^0_{\mathbf{n}, \alpha} + u_{\mathbf{n}, \alpha}$. The expression for the potential energy can be rewritten as
  \begin{equation}
U(.., \vec r_\mathbf{n}, ...) = \sum_{\mathbf{n}, \alpha} \frac{\partial U}{\partial r_{\mathbf{n}, \alpha}} ( r^0_{\mathbf{n}, \alpha} + u_{\mathbf{n}, \alpha}) + \frac{1}{2} \sum_{\mathbf{n}, \alpha} \sum_{\mathbf{l}, \beta} \frac{\partial^2 U}{\partial r_{\mathbf{n}, \alpha} \partial r_{\mathbf{l}, \beta}} ( r^0_{\mathbf{n}, \alpha} + u_{\mathbf{n}, \alpha}) ( r^0_{\mathbf{l}, \beta} + u_{\mathbf{l}, \beta}).
 \end{equation}
According to the Newton second law:
\begin{equation}
m \ddot u_{\mathbf{n}, \alpha} = - \frac{\partial U}{\partial u_{\mathbf{n}, \alpha}} = - \frac{\partial U}{\partial r_{\mathbf{n}, \alpha}} - \sum_{\mathbf{l}, \beta} \frac{\partial^2 U}{\partial r_{\mathbf{n}, \alpha} \partial r_{\mathbf{l}, \beta}} ( r^0_{\mathbf{n}, \alpha} + u_{\mathbf{n}, \alpha}).
 \end{equation}
By using the definition of the 0-displacement (2), we have 
\begin{equation}
m \ddot u_{\mathbf{n}, \alpha}  =  - \sum_{\mathbf{l}, \beta} \frac{\partial^2 U}{\partial r_{\mathbf{n}, \alpha} \partial r_{\mathbf{l}, \beta}}  u_{\mathbf{n}, \alpha}.
\end{equation}
Thus, with taking into account only the terms of the Tailor series containing derivatives of not higher than the second order, the 0-displacements are completely excluded from the equations of lattice vibrations. This result is the generalization of the known property of harmonic oscillator: the constant external field does not change its frequency.     

     Let us define the interface, at which the average 0-displacements are much less than the lattice constant, as ideal. Interaction of 0-displacements with lattice vibrations is revealed only if the terms of Tailor series with third and higher derivatives are taken into account. Thus, for an ideal interface the interaction of 0-displacements with lattice vibrations has the same order of magnitude as the interaction of phonons with each other and is revealed with taking into account the anharmonicity of vibrations. Hence, we can consider this interaction by means of perturbation theory.    

     Obviously, if the model of an ideal interface could describe properly the real interface, the latter should be sufficiently smooth, and the interaction between atoms on opposite sides of the interface essentially weaker than interaction between atoms of the same crystal. Atoms on opposite sides of the interface should not, in any case, form chemical bonds.

     We continue to consider lattice vibrations in the interfacial region using the ideal interface approximation and with no regard for anharmonicity. Ignoring the terms with derivatives higher than the third, we obtain the Hamiltonian function which is invariant with respect to displacements of the atoms of the left crystal by an amount equal to the lattice constant $a^L$, and also, of the atoms of the right crystal by the amount equal to the lattice constant $a^R$, along the axis $y$ or $z$. So, it is also invariant relative to any linear combinations of these displacements $n a^L + m a^R$ of each crystal. The symmetry group of this kind determines the specific properties of the lattice dynamics at the interface between two crystals.

\section{Equation for the quasi one-dimensional case}

Let’s consider first the more simple case where the phonon falls normally at the interface, that is, the wave vector components parallel to the interface  $q_y, q_z = 0$. 

    We define $u^L_{\mathbf{n}, \alpha}$ to be the displacement of the $\mathbf{n}$-th atom lying on the left side of the interface along the axis $\alpha$, and $u^R_{\mathbf{n'},\beta}$ the displacement of the $\mathbf{n'}$-th atom on the right side of the interface along the axis $\beta$. The Newton second law for the $\mathbf{l}$-th atom of the left crystal lying at the interface is then given by
 \begin{equation}
m \ddot u^L_{\mathbf{l}, \alpha}  =  - \sum_{\mathbf{n} \neq \mathbf{l}, \beta} \frac{\partial^2 U}{\partial u^L_{\mathbf{l}, \alpha} \partial u^L_{\mathbf{n}, \beta}}  u^L_{\mathbf{l}, \beta} - \sum_{\beta} \frac{\partial^2 U}{\partial u^L_{\mathbf{l}, \alpha} \partial u^L_{\mathbf{l}, \beta}} u^L_{\mathbf{l}, \beta} - \sum_{\mathbf{n'}, \beta} \frac{\partial^2 U}{\partial u^L_{\mathbf{l}, \alpha} \partial u^R_{\mathbf{n'}, \beta}}  u^R_{\mathbf{n'}, \beta} .
 \end{equation}

In the right part of Eq. (6) we separate out the term describing the interaction with the atoms of the same crystal as the atom under consideration, and the term describing the interaction with the atoms locating on the other side of the interface. To do this, we take into account that 

 \begin{equation}
\frac{\partial^2 U}{\partial u^L_{\mathbf{l}, \alpha} \partial u^L_{\mathbf{l}, \beta}}  =  - \sum_{\mathbf{n} \neq \mathbf{l}} \frac{\partial^2 U}{\partial u^L_{\mathbf{l}, \alpha} \partial u^L_{\mathbf{n}, \beta}} - \sum_{\mathbf{n'}} \frac{\partial^2 U}{\partial u^L_{\mathbf{l}, \alpha} \partial u^R_{\mathbf{n'}, \beta}},
 \end{equation}
as follows from the fact that the energy does not change if the atoms of both crystals are displaced by an equal distance in the same direction. Substitution of Eq. (7) in Eq. (6) gives
 \begin{equation}
m \ddot u^L_{\mathbf{l}, \alpha}  =  - \sum_{\mathbf{n} \neq \mathbf{l}, \beta} \frac{\partial^2 U}{\partial u^L_{\mathbf{l}, \alpha} \partial u^L_{\mathbf{n}, \beta}}  (u^L_{\mathbf{n}, \beta}-u^L_{\mathbf{l}, \beta}) - \sum_{\mathbf{n'}, \beta} \frac{\partial^2 U}{\partial u^L_{\mathbf{l}, \alpha} \partial u^R_{\mathbf{n'}, \beta}}  (u^R_{\mathbf{n'}, \beta}-u^L_{\mathbf{l}, \beta}) .
 \end{equation}

The first term in the right part of Eq. (8) describes the interaction of an atom with atoms of the same crystal, the second one relates to the interaction with atoms lying on the other side of the interface. 

     We seek a solution in the form of superposition of the incident, reflected and transmitted waves. We do not take into consideration the wave scattering at the interface so that the wave vector components of the reflected and transmitted waves, parallel to the interface, equal zero.  In this case, there are only three reflected and three transmitted waves with different polarization. Let us assume that from the left, at the interface falls the wave of unit amplitude and polarization $1$. Then, for an atom on the left side (not necessary at the interface itself) we obtain
 \begin{align}
 u^L_{\mathbf{n},\alpha} = \exp{(i \omega t)} \bigl( \exp{(- i q^L_1 a^L n_x)} \, e^L_{1\alpha} + A_1 \exp{(i q^L_1 a^L n_x)} \, e^L_{1\alpha} + \nonumber \\ + A_2 \exp{(i q^L_2 a^L n_x)} \, e^L_{2\alpha} + A_3 \exp{(i q^L_3 a^L n_x)} \, e^L_{3 \alpha} \bigr). 
 \end{align}

     Here indexes $1,2,3$ denote polarization, $A_{1,2,3}$ amplitudes of the reflected waves with different polarization, $q_{1,2,3}$ are $x$-components of the wave vectors. $\vec e^L_{1, 2, 3}$ stand for the polarization vectors, $e_{\alpha}$ for the component of the polarization vector in direction of the axis $\alpha$. At the interface $n_x = 0$, so for interfacial atoms each exponent in Eq. (9) becomes unity. 

The wave vector values $q_{1,2,3}$ are assumed to be known for the given frequency $\omega$, because the disperse relations for the waves in a simple cubic lattice are known \cite{Ans}. Near the interface, it also may be that the values $q^L_{2,3}$ are imaginary, that is, the oscillation is gradually damping  inward the crystal. It happens when the frequency of the incident wave is larger than the maximum frequency of oscillations in the reflected wave with the corresponding polarization \cite{Me}. Further on, we assume that the phonon falls at the interface from the left, and so the crystal from the side of which the phonon falls, will be called as "left" for brevity. 

     Similarly, for crystal atoms to the right of the interface we have
\begin{align}
 u^R_{\mathbf{n'},\alpha} = \exp{(i \omega t)} \bigl(B_1 \exp{(-i q^R_1 a^R n'_x)} \, e^R_{1 \alpha} + B_2 \exp{-i q^R_2 a^R n'_x)} \, e^R_{2 \alpha} + \nonumber \\ B_3 \exp{(-i q^R_3 a^R n'_x)} \, e^R_{3 \alpha} \bigr), 
 \end{align}
where $B_{1,2,3}$ are the amplitudes of the transmitted waves. The values $q^R_{1,2,3}$ can be imaginary (see the paper \cite{Me} and discussion in section $\mathbf{7}$). 

Equations (9) and (10) are the solutions of the Newton equations for the atoms lying off the interface, since $q_{1,2,3}$ and $\omega$ satisfy the dispersion relations for lattice vibrations, derived without regard for the boundary. The problem thus reduces to finding the $A_{1,2,3}, B_{1,2,3}$ which would satisfy the Newton equations for the interfacial atoms. For this purpose we substitute Eqs.(9) and (10) in Eq.(8) and try to simplify the expression obtained. For the term describing the interaction of an atom with atoms of the same crystal, we can proceed in a way similar to that in the paper of \cite{Young}. We divide both parts of Eq.(8) by $\exp{(i \omega t)}$, and then for the first term in the right part we have 
 \begin{align}
 \sum_{\mathbf{n} \neq \mathbf{l}, \beta} \frac{\partial^2 U}{\partial u^L_{\mathbf{l}, \alpha} \partial u^L_{\mathbf{n}, \beta}}  (u^L_{\mathbf{n}, \beta}-u^L_{\mathbf{l}, \beta}) = 
\nonumber \\  
=  \sum_{\mathbf{n} \neq \mathbf{n}, \beta} \frac{\partial^2 U}{\partial u^L_{\mathbf{n}, \alpha} \partial u^L_{\mathbf{n}, \beta}}  \Bigl( 
(1 -\exp{(- i q^L_1 a^L n_x)}) e^L_{1 \beta} + \sum_{j}  A_j (1-\exp{(i q^L_j a^L n_x)}) e^L_{j \beta} \Bigr) = 
\nonumber \\  
=   \sum_{\mathbf{n} \neq l, \beta} \frac{\partial^2 U}{\partial u^L_{\mathbf{l}, \alpha} \partial u^L_{\mathbf{n}, \beta}} (1-\exp{(- i q^L_1 a^L n_x)}) e^L_{1 \beta} + 
\nonumber \\  
+ \sum_{j}  A_j \sum_{\mathbf{n} \neq \mathbf{l}, \beta} \frac{\partial^2 U}{\partial u^L_{\mathbf{l}, \alpha} \partial u^L_{\mathbf{n}, \beta}} (1-\exp{(- i q^L_j a^L n_x)}) e^L_{j \beta}.
 \end{align}
We take into account that the frequency and the polarization vector can be expressed in terms of the dynamic matrix \cite{Ans}
 \begin{equation}
\omega^2 e_{j \alpha} = \frac{1}{m} \sum_{\mathbf{l} \neq \mathbf{n}, \beta} \frac{\partial^2 U}{\partial u^L_{\mathbf{n}, \alpha} \partial u^L_{\mathbf{l} , \beta}} (1-\exp{(- i q_j a^L (n_x-l_x)}) e_{j \beta} =  \sum_{\beta} D_{\alpha \beta} e_{j \beta} (q_j).
 \end{equation}

In Eq.(11) the summation is over the atoms of the same crystal only. To the right of the interfacial atom of the left crystal, $n_x = 0$, there are no more atoms of that crystal, $n_x \leq 0$, that can be formally taken into consideration in the form of the condition $l_x \leq n_x$. We introduce the notation (see Fig. 2)
 \begin{align}
 D^{\times}_{\alpha \beta}(q_j) =  \frac{1}{m} \sum_{\substack{\mathbf{l} \neq \mathbf{n}, \beta \\ l_x \leq n_x}} \frac{\partial^2 U}{\partial u^L_{\mathbf{n}, \alpha} \partial u^L_{\mathbf{l}, \beta}} (1-\exp{(- i q^L_j a^L l)}) .
 \end{align}

which is for the dynamic matrix describing the atom lying at the interface. Here we can ignore the fact that $n_x \leq 0$, since for an interfacial atom it was taken into account in the expression $l_x \leq n_x$. So we can operate with Eq.(13) formally.

     Denote the difference of the two dynamic matrices (one for an atom in the depth of the crystal and the other for an atom at the interface) as 
 \begin{equation}
 ^{\otimes}D_{\alpha \beta}(q_j) = D_{\alpha \beta}(q_j) - D^{\times}_{\alpha \beta}(q_j) 
= \frac{1}{m} \sum_{\substack{\mathbf{l} \neq \mathbf{n}, \beta \\ l_x > n_x}} \frac{\partial^2 U}{\partial u^L_{\mathbf{n}, \alpha} \partial u^L_{\mathbf{l}, \beta}} (1-\exp{(- i q^L_j a^L l_x)}) .
 \end{equation}

We substitute Eqs.(12), (13) in Eq.(8) and transfer the first term of the right part of Eq.(8) to the left. The expression thus obtained can be written in new notations in the form

\begin{equation}
\sum_{j}  A_j \sum_{\beta} \, ^{\otimes}D_{\alpha \beta}(- q_j) e^L_{j \beta} +  \, ^{\otimes}D_{\alpha \beta}(q_1) e^L_{1 \beta} = - \sum_{\mathbf{n'}, \beta} \frac{\partial^2 U}{\partial u^L_{\mathbf{n}, \alpha} \partial u^R_{\mathbf{n'}, \beta}}  (u^R_{\mathbf{n'}, \beta}-u^L_{\mathbf{n}, \beta}).
 \end{equation}

Now we rearrange the right part of Eq.(15) by using the relation 
 \begin{equation}
K_{\mathbf{n}, \alpha \beta} = \sum_{\mathbf{n'}} \frac{\partial^2 U}{\partial u^L_{\mathbf{n}, \alpha} \partial u^R_{\mathbf{n'}, \beta}} 
\end{equation}
which is the matrix of the interfacial interaction, describing the interaction of an atom with atoms lying on the opposite side of the interface, where $n_x = 0$ is supposed. Unlike the conventional dynamic matrix, the matrix of this kind depends on the number of the interfacial atom, because each atom at the interface is differently located relative to atoms on the opposite side of the interface. 

\begin{figure}
\centering
\includegraphics[width=0.6\textwidth]{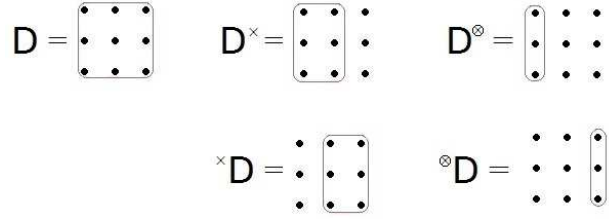}
\caption{Schematically displayed is the interaction between atoms which is taken into account in a given dynamic matrix.}
\end{figure}

Substituting Eqs.(10), (16) into Eq.(15), we come to 
 \begin{equation}
\sum_{j}  A_j \sum_{\beta} \, ^{\otimes}D_{\alpha \beta}(q_j) e^L_{j \beta} +  \, ^{\otimes}D_{\alpha \beta}(-q_1) e^L_{1 \beta} = - \sum_{\beta} K_{\mathbf{n}, \alpha \beta} ( e^L_{1 \beta} + \sum_{j}  (A_j  e^L_{j \beta} - B_j  e^R_{j \beta}).
 \end{equation}

Then we sum over all $n$ in Eq.(17) and divide the result by a total number of atoms $N$ at the interface on the left. The left part of the expression is unchanged for it is independent of the number $n$ of the atom. In the right part of the expression we obtain

\begin{equation}
K_{\alpha \beta}^{L} = \frac{1}{N} \sum_{\mathbf{n}}  K_{\mathbf{n}, \alpha \beta}.
\end{equation}

This is the averaged matrix describing the interaction of atoms across the interface. By performing such an averaging, we do not take into account that the atoms nearest to the interface differ in their position relative to atoms on the opposite side of the interface, and, hence, interact differently with them. Taking account of this difference involves the appearance of scattering, which calls for mathematical technique presented below in Sec. $\mathbf{5}$. In this section we proceed ignoring the scattering. 

     With the simplifications made above ($q_{y,z} = 0$ for all reflected and transmitted waves), all atoms with the same coordinate along the axis $x$ are oscillating in phase. In fact, performing such an averaging, we do not consider oscillations of an individual atom, but oscillations of the whole of the plane near the interface of the crystal. The matrix $K_{\alpha \beta}^{L}$ describes forces acting on the crystal plane on the left of the interface from the side of the crystal plane lying on the right of the interface. The problem in this case becomes quasi-one-dimensional.

     Let the relation 
 \begin{equation}
M_{\alpha j}^{L, \pm}  = \sum_{\beta} \bigl(\,  ^{\otimes} D_{\alpha \beta}(\pm q_j) + K_{\alpha \beta} \bigr) e^L_{j \beta}
\end{equation}
be the second dynamical matrix. It expresses the forces acting on atoms near the interface in terms of amplitudes of waves differing in polarization on the left side of the interface. That to write the analogous expression for the forces acting on atoms on the other side of the interface, we introduce  

\begin{equation}
I_{\alpha j}^{L} = \sum_{\beta} K_{\alpha \beta}  e^R_{j \beta}.
\end{equation}

With new notations, Eq.(17) can be rewritten in the form 
\begin{equation}
M_{\alpha j}^{L, +} A_j + I_{\alpha j}^L B_j =M_{\alpha j}^{L, -}.
\end{equation}
where the summation is supposed to be performed over the repetitive index $j$ standing for the wave polarization.

     By writing the Newton second law for atoms on the right of the interface, one can rearrange the obtained expression in a similar way. A distinction is in the fact that the averaged matrix of the interfacial interaction is defined as 

\begin{equation}
K_{\beta \alpha}^{R} = \frac{1}{N'} \sum_{\mathbf{n'}}   \sum_{\mathbf{n}} \frac{\partial^2 U}{\partial u^L_{\mathbf{n}, \alpha} \partial u^R_{\mathbf{n'}, \beta}} = \frac{N}{N'} K_{\alpha \beta}^{L},
\end{equation}

where $N'$ is a number of atoms nearest to the interface in the crystal on the right. The amounts of atoms near the interface are related as inverse squares of the lattice constants. Thus we obtain
\begin{equation}
K_{\beta \alpha}^{R} = \Bigl(\frac{a^L}{a^R} \Bigr)^2 K_{\alpha \beta}^{L},
\end{equation}

which represents the fact that the force acting on the right crystal from the side of the left one is equal, in magnitude, to the force acting on the left crystal from the side of the right one. Finally, we have
\begin{align}
M_{\alpha j}^{L, +} A_j + I_{\alpha j}^L B_j &= M_{\alpha 1}^{L, -} \nonumber \\
M_{\alpha j}^{R, -} B_j + I_{\alpha j}^R A_j &= 0.
\end{align}

This is the sought-for system of equations.

\section{Theorem on the interfacial interaction and the exact solution in the quasi-one-dimensional case}

Let us study the matrix of interfacial interaction, $K_{\alpha, \beta}$, using the fact that the interaction energy of two crystals is invariant in respect to displacements by any linear combination of the lattice vectors. We consider first the general case when the lattice parameters are incommensurate, that is, their ratio is irrational, $a^L/a^R \notin \mathbb{Q}$.  In this case, linear combinations of the lattice vectors and, hence, a set of rearrangements converting the system into itself, is dense on the plane. It is reasonable to require that the energy be continuous as the function of the relative displacement of the crystals. Since the energy is constant on the dense set and, besides, is continuous, it is constant. 

     To be specific, we will demonstrate that $K_{y, y} = 0$. We displace the first crystal by a small distance $Y$ along the axis $y$. The energy in a new state will, on the one hand, be equal to the initial one, and on the other, be different an amount of the work done by the force of interaction between crystals:

 \begin{equation}
E = E +\int_0^Y K_{y y} y dy \Rightarrow K_{y y} Y^2/2 = 0 \Rightarrow K_{y,y} = 0.
 \end{equation}
For the other seven components of the matrix $K_{\alpha \beta}$ (except for $K_{x x}$), the proof can be performed in a similar way. Thus, we conclude that, with initial assumptions, the only nonzero component of the matrix of interfacial interaction is $K_{x, x}$.

     We consider now a case of resonance where $a^L/a^R \in \mathbb{Q}$. It is obvious that the above reasoning is inapplicable. Let $a^L/a^R = p/q$ and $p/q<1$. The interaction energy of two crystals will be described by a periodical function of their relative displacement with a period $q a^L$. We expand it and $K_{\alpha \beta}$ into the Fourier series and substitute the result in Eq.(21). The first, largest term of the series turns out to be 

\begin{equation}
K_{\alpha \beta 1}^{R} = \frac{1}{q^2 (a^L)^2} U_1
\end{equation}
We drop the other terms and obtain the following assessment for the case of equal potentials of the interatomic interaction: 
\begin{align}
 K_{\alpha \beta } \sim 1/q^2,& \ a^L/a^R = p/q  \nonumber \\
K_{\alpha \beta } = 0,&  \ a^L/a^R \notin \mathbb{Q}.
\end{align}
Such a function experiences a discontinuity in all rational points, which is impossible from the physical point of view. The paradox can be resolved from considerations that this function was obtained ignoring the interaction with 0-displacements which are large near the resonance. Accounting for the anharmonic coupling with 0-displacements leads to smearing of spikes in rational points. As a result, the actual matrix of the interfacial interaction will be given by a smooth envelope of the function (26).

Hence, we come to the following qualitative result: the atomic oscillations perpendicular to the plane of the interface interact intensively with the atomic oscillations on the opposite side of the interface only in the case when a ratio of the lattice constants is close to a rational number with a small denominator. Otherwise, this interaction is weak. 

     This result can be illustrated by an interesting analogy in the nebular mechanics, where the close commensurability of revolution periods of planets, i.e., the proximity of a ratio of periods to a rational number with a small denominator, leads to the strong long-period perturbation \cite{Arn}.

     Using the predetermined theorem on the matrix of the interfacial interaction, one can obtain the exact solution in the quasi-one-dimensional nonresonance case. We write explicitly the expressions for the elements of the matrix $ ^{\otimes}D_{\alpha \beta}(q_j)$:
\begin{align}
^{\otimes}D_{x x}(q_j) = \frac{\beta^L_1 + 2 \beta^L_2}{m}(1 - e^{iq_x a^L})  \nonumber \\
^{\otimes}D_{y y}(q_j) = \, ^{\otimes}D_{z z}(q_j) = \frac{ \beta^L_2}{m}(1 - e^{iq_x a^L}) 
 \nonumber \\
^{\otimes}D_{x y}(q_j) = \, ^{\otimes}D_{x z}(q_j)  = \, ^{\otimes}D_{y z}(q_j) = 0.
\end{align}
The remaining three components are also equal zero since the matrix is symmetrical. 

For the case of the normal incidence of the wave at the interface, we have 
\begin{align}
e^L_1 = (1,0,0)^T \nonumber \\
e^L_2 = (0,1,0)^T \nonumber \\
e^L_3 = (0,0,1)^T
\end{align}
Notations: $K^L_{x x} = \beta^L$.
Let a longitudinally polarized wave is incident at the interface. Substituting Eqs.(27), (28) in Eq.(18), we get
 \begin{equation}
M_{x, 1}^{L, \pm}  =\frac{\beta^L_1 + 2 \beta^L_2}{m}(1 - e^{\pm iq_x a^L}) +\frac{\beta^L}{m}
\end{equation}
In addition, one can easily see that all nondiagonal components of the matrix $M$ equal zero. Hence, when the incident wave is longitudinally polarized, the reflected and transmitted waves have also the longitudinal polarization. The system of six linear equations (23) transforms into the simple system of two linear equations:
\begin{align}
M_{x, 1}^{L, +} A_1 - \beta^L B_1 = - M_{x, 1}^{L, -} \nonumber \\
M_{x, 1}^{R, +} B_1 - \beta^R (1+A_1) = 0 .
\end{align}
Whence it follows 
\begin{align}
 A_1 = \frac {\beta^L \beta^R -  M_{x, 1}^{L, -} M_{x, 1}^{R, +}}{M_{x, 1}^{L, +}M_{x, 1}^{R, +} - \beta^L \beta^R} \nonumber \\
B_1 = \frac{\beta^R}{M_{x, 1}^{R, +}} \frac{M_{x, 1}^{L, +} - M_{x, 1}^{L, 1} } {M_{x, 1}^{L, +} - \beta^L \beta^R} .
\end{align}
This solution coincides with the result obtained in the paper \cite{Zh} for a one-dimensional chain if one rewrite it in the above notations. In the one-dimensional case, the 3- by-3 matrices are changed by a single number, and $\beta^L = \beta^R$. Hence, the results obtained for a one-dimensional chain would be applicable in the three-dimensional case. 

     Further on we consider the case where the wave with transverse polarization is incident at the interface. For the sake of definiteness, we assume that the oscillations occur along the axis $y$. Since according to the theorem on the matrix of interfacial interaction we have $K_{y y} = 0$, the corresponding equations for the transmission coefficient have the form
\begin{align}
M_{y, 2}^{L, +} A_2  = - M_{y, 1}^{L, -} \nonumber \\
M_{y, 2}^{R, +} B_2 = 0 ,
\end{align}
whence $B = 0$, $|A| = 1$.

     Actually, however, $K_{y y}$ is small but not equals zero identically, because the 0-displacements, though being small, are nonzero. It was shown in the paper \cite{Zh} that at small frequency of the incident wave, $\omega \rightarrow 0$, the transmission coefficient is independent of the interaction force of atoms at the interface, but determined by acoustic impedances of media on opposite sides of the interface, in other words, the transmission coefficient can be found from the elasticity theory. However, when the interaction force of interfacial atoms is weak, the transmission coefficient of phonons decreases fast with growing frequency. Since the heat transport is carried out by phonons of all frequencies, we can neglect the contribution to the energy transport across the interface from atomic oscillations parallel to the interface.

\section{Fourier transform of the matrix of interfacial interaction}

In order to study the phonon transmission across the interface between two crystals in more general case when a phonon is incident at arbitrary angle, and therewith take into consideration the scattering, the special mathematical apparatus is required which is presented below. For simplicity we consider a one-dimensional case, becouse the generalization to more dimensions is trivial. 

     Let we have an infinite one-dimensional chain of atoms spaced $a^L$ apart. The atoms of the chain are numbered from minus infinity to plus infinity, and the axis $x$ is located so that the zero-th atom has the coordinate $x=0$. We place the chain into the external potential $\Phi(x)$ with a period $a^R$, $\Phi(x+a^R) = \Phi(x)$.

     Let’s expand the $\Phi(x)$ into the Fourier series:
 \begin{align}
\Phi_k =& \frac{1}{2 \pi a^R} \int \limits_0^{a^R} \Phi(x) e^{-2 \pi ikx/ a^R}\, dx   \nonumber \\
\Phi(x) =& \sum_{k=-\infty}^{+\infty} \Phi_k\, e^{2 \pi ikx/ a^R}
\end{align}

Then the potential energy of $n$-th atom, $\Phi_n$, will equal $\Phi(n a^L)$, or
\begin{equation}
\Phi_n = \sum_{k=-\infty}^{+\infty} \Phi_k\, e^{2 \pi ikn (a^L/ a^R)}
\end{equation}
One can demonstrate that the inverse rearrangement is also valid:
\begin{equation}
\Phi_k = \lim_{N \rightarrow \infty} \frac{1}{2N} \sum_{n=-N}^{N} \Phi_n\, e^{-2 \pi ikn (a^L/ a^R)}
\end{equation}

Actually, 
\begin{align}
 \lim_{N \rightarrow \infty} \frac{1}{2N} \sum_{n=-N}^{N} \Phi_n\, e^{-2 \pi ikn (a^L/ a^R)}  = \nonumber \\ 
= \lim_{N \rightarrow \infty} \frac{1}{2N} \sum_{n=-N}^{N} \Bigl( \sum_{k'=-\infty}^{+\infty} \Phi_{k'}\, e^{2 \pi ik'n (a^L/ a^R)} \Bigr)\, e^{-2 \pi ikn (a^L/ a^R)} =
 \nonumber \\
= \sum_{k'=-\infty}^{+\infty} \Phi_{k'} \lim_{N \rightarrow \infty} \frac{1}{2N} \sum_{n=-N}^{N} e^{2 \pi i(k'-k)n (a^L/ a^R)} = \sum_{k'=-\infty}^{+\infty} \Phi_{k'} \delta_{k' k} = \Phi_k
\end{align}

Thus, we can “forget” the initial function $\Phi(x)$ and consider only the discrete sets of values $\Phi_n$ and $\Phi_k$  which are expressed in terms of each other. 

     For what follows, Eq.(35) can be conveniently rewritten so that it would correspond to the expansion in exponents with the wave vectors $q_k$ lying in the first Brillouin zone, i.e., $q_k \in (-\pi/a^L, \pi/a^L)$. To do so, we introduce  
\begin{equation}
q'_k = \frac{2 \pi}{a^L} \left\{ k \frac{a^L}{a^R} \right\},
\end{equation}
where the curly brackets $\{ ... \}$ denote the fractional part. Then we have
\begin{equation}
\Phi_n = \sum_{k=-\infty}^{+\infty} \Phi_k\, e^{ i q'_k a^L n},
\end{equation}
Evidently, this expression is equivalent to Eq.(35), but $q'_k \in (0, 2 \pi/a^L)$. That to obtain the desired range of values of the wave vector, we have to take 
\begin{equation}
q_k = \frac{2 \pi}{a^L} \left( \left\{ k \frac{a^L}{a^R} + \frac{1}{2}\right\} - \frac{1}{2} \right),
\end{equation}
and, finally, we obtain 
\begin{equation}
\Phi_n = \sum_{k=-\infty}^{+\infty} \Phi_k\, e^{ i q_k a^L n}.
\end{equation}

Eq.(35) can be applied to the matrix $K_{\mathbf{n}, \alpha \beta}$ and that to use it for description of the phonon transmission across the interface of two crystal we have to rearrange it. Really, 
\begin{equation}
K_{\mathbf{n}, \alpha \beta} = \frac{\partial}{\partial u^L_{\mathbf{n}, \alpha} } \sum_{\mathbf{n'}} \frac{\partial U}{ \partial u^R_{\mathbf{n'}, \beta}}.
\end{equation}
The function
\begin{equation}
\Phi =  \sum_{\mathbf{n'}} \frac{\partial U}{ \partial u^R_{\mathbf{n'}, \beta}},
\end{equation}
is periodical (the period $a^R$), since the potential produced by atoms of the right crystal for atoms of the left crystal is periodical. Hence, we can introduce 
\begin{align}
 K_{\mathbf{k}, \alpha \beta} = \lim_{N \rightarrow \infty} \frac{1}{N} \sum_{\mathbf{n}} K_{\mathbf{n}, \alpha \beta}\, e^{2 \pi i \mathbf{k}\mathbf{n} (a^L/ a^R)} 
\nonumber \\
K_{\mathbf{n}, \alpha \beta} = \sum_{\mathbf{k}=-\infty}^{+\infty} K_{\mathbf{k}, \alpha \beta}\, e^{2 \pi i \mathbf{k}\mathbf{n} (a^L/ a^R)},
\end{align}
where $\mathbf{k} = (0, k_y, k_z)$.

\section{Equation and the theorem on the interfacial interaction in the general case}

     Let the wave of the unit amplitude, polarization $1$ and wave vector $\mathbf{q}_1$ falls at the interface. Notations: $\mathbf{q}_{||} = (0, q_y, q_z)$ are the wave vector components parallel to the interface, and $\mathbf{n}_{||} = (0, n_y, n_z)$ is the atom number along the axes $y, z$ parallel to the plane of the interface. We seek the solution for the left side in the form of superposition of the incident and reflected waves:

\begin{align}
 u^L_{\mathbf{n},\alpha} = \exp{(i \omega t)} \Bigl( \exp{(- i q_{1,x}^L n_x a^L + i \mathbf{q}_{||}^L \mathbf{n}_{||} a^L)}  e^L_{1 \alpha} + \nonumber \\ + \sum_{j, \mathbf{k}}  A_{\mathbf{k}, j}  \exp{( i q_{j,x}^L n_x a^L + i (\mathbf{q}_{||} + 2 \pi i \mathbf{k}/a^R) \mathbf{n}_{||} a^L )} \, e^L_{\mathbf{k}, j \alpha}  \Bigr), 
 \end{align}
And for the right side in the form of superposition of the transmitted waves:\begin{align}
 u^R_{\mathbf{n'}, \alpha} = \exp{(i \omega t)} \sum_{j, \mathbf{k'}} 
 B_{\mathbf{k'}, j }  \exp{( i q_{j,x}^L n'_x a^R + i (\mathbf{q}_{||}+ 2 \pi i \mathbf{k'}/a^L) \mathbf{n'}_{||} a^R )} \, e^L_{\mathbf{k}, j \alpha} .
 \end{align}

We substitute Eqs.(45), (46) into Eq.(8). On rearrangement of the left part and of the first term in the right part we have
 \begin{align}
\exp{ i (\mathbf{q}_{||} \mathbf{n}_{||} a^L )} \sum_{\mathbf{k},j} \exp{ ( 2 \pi i \mathbf{k} \mathbf{n}_{||} a^L / a^R)}  A_{\mathbf{k},j} \sum_{\beta} \, ^{\otimes}D_{\alpha \beta}(-\mathbf{q}_{j,x}^L +\mathbf{q}_{||} + 2 \pi i \mathbf{k}/a^R) e^L_{j, \beta} +
\nonumber \\
+ \exp{ i (\mathbf{q}_{||} \mathbf{n}_{||} a^L )}  \, ^{\otimes}D_{\alpha \beta}(\mathbf{q}) e^L_{1, \beta} 
= - \sum_{\mathbf{n'}, \beta} \frac{\partial^2 U}{\partial u^L_{\mathbf{n}, \alpha} \partial u^R_{\mathbf{n'}, \beta}}  (u^R_{\mathbf{n'}, \beta}-u^L_{\mathbf{n}, \beta}).
 \end{align}

The first term of the right part of Eq.(47) can be written as
\begin{equation}
\sum_{\mathbf{n'}, \beta} \frac{\partial^2 U}{\partial u^L_{\mathbf{n}, \alpha} \partial u^R_{\mathbf{n'}, \beta}} u^L_{\mathbf{n}, \beta} = \sum_{ \beta} K^L_{\mathbf{n}, \alpha \beta} u^L_{\mathbf{n}, \beta},
\end{equation}
since $ u^L_{\mathbf{n}, \beta}$ is independent of $\mathbf{n'}$ and could be taken out of the sum over $\mathbf{n'}$.

In the second term of the right part of Eq.(47) we make the following rearrangement:\begin{equation}
\sum_{\mathbf{n'}, \beta} \frac{\partial^2 U}{\partial u^L_{\mathbf{n}, \alpha} \partial u^R_{\mathbf{n'}, \beta}} u^R_{\mathbf{n'}, \beta} = \exp{( i \mathbf{q}_{||} \mathbf{n}_{||} a^L)}  \sum_{\mathbf{n'}, \beta} \frac{\partial^2 U}{\partial u^L_{\mathbf{n}, \alpha} \partial u^R_{\mathbf{n'}, \beta}}  \exp{( - i \mathbf{q}_{||} \mathbf{n}_{||} a^L)} u^R_{\mathbf{n'}, \beta}.
\end{equation}

     Let’s divide Eq.(47) by $\exp{ i (\mathbf{q}_{||} \mathbf{n}_{||} a^L )}$ and multiply by $\exp{(2 \pi i \mathbf{l}/a^R) \mathbf{n}_{||} a^L )}$.  Then we sum over $\mathbf{n}$ so that the total number of additives would equal $N$, and divide by $N$. The left part acquires the form
\begin{equation}
\sum_{\mathbf{k},j}  A_{\mathbf{l - k}, j} \sum_{\beta } K_{\mathbf{k}, \alpha \beta}  e^L_{\mathbf{k}, j \beta} +\delta_{\mathbf{l} \mathbf{0}}  \sum_{\beta }  K_{\mathbf{0}, \alpha \beta}  e^L_{\mathbf{0}, 1 \beta}   .
\end{equation}

For the second term we introduce
\begin{equation}
K_{\mathbf{n}, \alpha \beta} (\mathbf{q}_{||}) = \sum_{\mathbf{n'}_{||}, \beta} \frac{\partial^2 U}{\partial u^L_{\mathbf{n}, \alpha} \partial u^R_{\mathbf{n'}, \beta}} e^{ i \mathbf{q}_{||} ( \mathbf{n'}_{||} a^R - \mathbf{n}_{||} a^L )},
\end{equation}

\begin{equation}
K^L_{\mathbf{k} \mathbf{k'}, \alpha \beta} (\mathbf{q}_{||}) = \sum_{\mathbf{n'}_{||} \mathbf{n}_{||}, \beta} \frac{\partial^2 U}{\partial u^L_{\mathbf{n}, \alpha} \partial u^R_{\mathbf{n'}, \beta}} 
e^{ i \mathbf{q}_{||} ( \mathbf{n'}_{||} a^R - \mathbf{n}_{||} a^L )} 
e^{  2 \pi i (\mathbf{k}_{||} \mathbf{n}_{||} a^L/a^R + \mathbf{k'}_{||} \mathbf{n'}_{||} a^R/a^L    )    }
\end{equation}

In these notations the second term of the right part of Eq.(47), after the rearrangement, can be written as

\begin{equation}
\sum_{j, \mathbf{k'}}  B_{\mathbf{k'}, j} \sum_{\beta } K^L_{\mathbf{l} \mathbf{k'}, \alpha \beta} (\mathbf{q}_{||}) e^R_{\mathbf{k'}, j \beta}.
\end{equation}

Similarly, as it has been made in section $\mathbf{3}$, we introduce the second dynamic matrices:

the matrix describing the interaction of the waves with the same $\mathbf{q}_{||}$:
 \begin{equation}
M_{\mathbf{k}, \alpha j}^{L, \pm}  = \sum_{\beta} \bigl(\,  ^{\otimes} D_{\alpha \beta}(\pm \mathbf{q}_{j,x}^L +\mathbf{q}_{||} + 2 \pi i \mathbf{k}/a^R) + K_{0, \alpha \beta} \bigr) e^L_{\mathbf{k}, j \beta} ,
\end{equation}
the matrix describing the interaction with the oscillations on the opposite side of the interface:\begin{equation}
I_{\mathbf{k}, \alpha j}^{L}  (\mathbf{q}_{||}) =  \sum_{\beta } K^L_{\mathbf{k} \mathbf{k'}, \alpha \beta} (\mathbf{q}_{||}) e^R_{\mathbf{k'}, j \beta}.
\end{equation}
and the matrix describing the interaction with the oscillations of the same crystal, but with another $\mathbf{q}_{||}$:
 \begin{equation}
S_{\mathbf{k}, \alpha j}^{L, \pm}  = \sum_{\beta } K_{\mathbf{k}, \alpha \beta}  e^L_{\mathbf{k}, j \beta}.
\end{equation}

In these notations Eq.(47) will be rewritten as

\begin{equation}
M_{\mathbf{l}, \alpha j }^{L, +} A_{\mathbf{l}, j } +  \sum_{\mathbf{k}} S_{\mathbf{k}, \alpha j}^L A_{\mathbf{l-k}, j }  + \sum_{\mathbf{k'} } I_{\mathbf{k'}, \alpha j}^L  (\mathbf{q}_{||}) B_{\mathbf{k'}, j } = M_{\mathbf{0}, \alpha 1}^{L, -} \delta_{\mathbf{l} \mathbf{0}}
\end{equation}

For the right crystal a similar approach gives:\begin{equation}
M_{\mathbf{l'}, \alpha j }^{R, -} B_{\mathbf{l}, j } +  \sum_{\mathbf{k'}} S^R_{\mathbf{k'}, \alpha j} B_{ j \mathbf{l'-k'}}  + \sum_{\mathbf{k}} I^R_{\mathbf{k}, \alpha j} A_{\mathbf{k}, j } = 0
\end{equation}

Eqs.(59) and (60) taken together form the complete system of equations describing the transmission, reflection and scattering of phonons at the interface between two crystals.

     Thus, it is seen that, on the one hand, even in the model of an ideal interface without defects and roughness, there appears the scattering due to the mismatch of the crystal lattices. On the other hand, the assumption in DMM that a phonon incident on the interface “forgets” its initial direction is, generally saying, incorrect. The scattering reveals quite determined structure: the wave vector of the transmitted wave, parallel to the interface, differs from the wave vector of the incident wave by an integer number of the vectors of the reciprocal lattice of the left crystal, while the wave vector, parallel to the interface, of the reflected wave differs from the wave vector of the incident wave by an integer number of the vectors of the reciprocal lattice of the right crystal. That the vectors of the scattered phonons lie in the first Brillouin zone it is remained to rearrange them using Eq.(40). 

      It is impossible to solve the infinite system of Eqs.(59) and (60) analytically. However, if the function $\Phi$, in terms of which the matrix $K_{\mathbf{n}, \alpha \beta}$ is determined in Eq.(42), is rather smooth, the coefficients of its expansion into the Fourier series fast decreases and along with them the matrices $K_{\mathbf{k}, \alpha \beta}$ describing the scattering also do. In this case one can choose the definite number of the terms from the system (59, 60) and solve it numerically. In the case when the matrices $K_{\mathbf{k}, \alpha \beta}$ are large even at large values of $\mathbf{k}$, one can consider the assumption of the DMM as being valid. 

     In the case when the scattering is small, one can alternatively use the perturbation theory, by taking the solution of the system of six equations describing the phonon transmission without scattering as the unperturbed one:
 \begin{align}
M_{\mathbf{0}, \alpha j }^{L, +} A_{\mathbf{0}, j } +  I_{\mathbf{0}, \alpha j}^L  (\mathbf{q}_{||}) B_{\mathbf{0}, j } = M_{\mathbf{0}, \alpha 1}^{L, -} 
\nonumber \\
M_{\mathbf{0}, \alpha j }^{R, -} B_{\mathbf{0}, j } +  I^R_{\mathbf{0}, \alpha j}  (\mathbf{q}_{||}) A_{\mathbf{0}, j } = 0.
 \end{align}
This system describes the transmission of phonons across the interface with no regard for scattering, i.e. the refraction of phonons at the interface of two crystals. 

Unfortunately, even this system cannot be solved analytically, since it is impossible to inverse the expression (12) and express the wave vector component normal to the interface, $q_x$, in terms of frequency and two other components (except for the case of high symmetry, $q_y = q_z = 0$, considered in section $\mathbf{4}$). Such an inversion can be performed for the interface of face-centered cubic lattices and was made in the work \cite{Young}. In the present paper we deal with the model of simple cubic lattices because just this model ensured the simplest qualitative study of the impact of the crystal lattice mismatch on the transmission of phonons across the interface. 

     To facilitate the numerical calculation of the problem, it is reasonable to invoke the theorem on the matrix of the interfacial interaction in the general. Let’s expand the function $\Phi$ into the Fourier series:
\begin{equation}
\Phi (x) =  \sum_{\mathbf{k}=-\infty}^{+\infty}  \Phi_k\, e^{2 \pi i\mathbf{k}x/ a^R}.
\end{equation}
On substitution into Eq. (42), we find that the matrix $ K^L_{\mathbf{k}, \alpha \beta}$ is expressed in terms of the components of $\Phi$ as
\begin{equation}
K^L_{\mathbf{k}, \alpha \beta} = k_\alpha \Phi_k.
\end{equation}
Thus, $K_{(0, k_z), y \beta} = 0$ and $K_{(k_y, 0), z \beta} = 0$. We use the fact that according to Eq.(23), $ K_{\beta \alpha}^{R} = (a^L/a^R)^2 K_{\alpha \beta}^{L}$ and that for $K_{\beta \alpha}^{R}$ the expression analogous to Eq.(42) is valid. We obtain
 \begin{equation}
K^L_{0, \alpha \beta} = \beta \delta_{\alpha x} \delta_{\beta x}.
\end{equation} 
Thus, the atomic oscillations parallel to the interface do not contribute to the transmission of phonons without scattering at the interface.

\section{Refraction of phonons at the crystal interface}

Eq.(60) admits the analytical solution in the case when a phonon falls at the interface at a small angle. Then, by taking $\mathbf{q}_{||}$ as a small parameter and applying the perturbation theory, we can find the correction to the amplitudes of the reflection and transmission, which is proportional to $q^2_{||}$. This solution is extraordinary cumbersome and so we omit it here. 

 \begin{figure}
\centering
\includegraphics[width=0.6\textwidth]{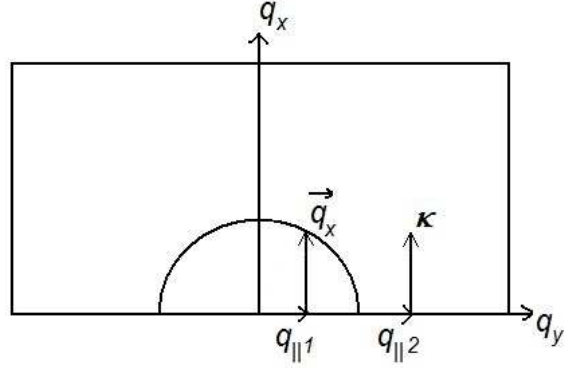}
\caption{ A plane of constant frequency in the Brillouin zone. It is shown that a phonon with the component of the wave vector parallel to the interface, $q_{||1}$ passes through it. When an incident phonon has the component of the wave vector parallel to the interface $q_{||2}$, on the opposite side there appears an inward-damping oscillation with the damping coefficient $\kappa$.}
\end{figure}

Instead we will give a qualitative description of refraction of phonons at the crystal interface, for any angle of incidence. To do so, we display the surface of the constant frequency $\omega$ in the Brillouin zone (Fig.3) and make a projection of this surface onto the plane $(q_y, q_z)$. If the component $\mathbf{q}_{||}$ of the incident phonon is lying in limits of this projection, it is easy to see that for the transmitted phonon the wave vector component normal to the interface, $q_x$, is equal to the $q_x$ of that point in the Brillouin zone which was projected into the point $\mathbf{q}_{||}$.  

     If the component $\mathbf{q}_{||}$ of the incident phonon lies beyond the projection, the transmitted phonon does not appear, because there is no the $q_x$ such that the phonon with the wave vector $(q_x, q_{||y}, q_{||z})$ has the frequency $\omega$. In this case there occurs the oscillation directed into the depth from the interface, with the damping coefficient $\kappa$ and wave vector parallel to the interface $\mathbf{q}_{||}$. The specific cases of such oscillations are well known Rayleigh waves \cite{Ld}. Just as there are three branches of oscillations for the given $\omega, \mathbf{q}_{||}$, one should perform the above procedure for each of them. 

     If the frequency of the incident wave exceeds the maximum possible in the given branch, there occurs in the crystal the inward-damping oscillation at any value of $\mathbf{q}_{||}$. The oscillations of this kind were predicted in the model of a one-dimensional chain in paper \cite{Me}. Probably, it is such oscillations that were found in numerical simulations in work \cite{Nuo}. 

 \begin{figure}
\centering
\includegraphics[width=0.6\textwidth]{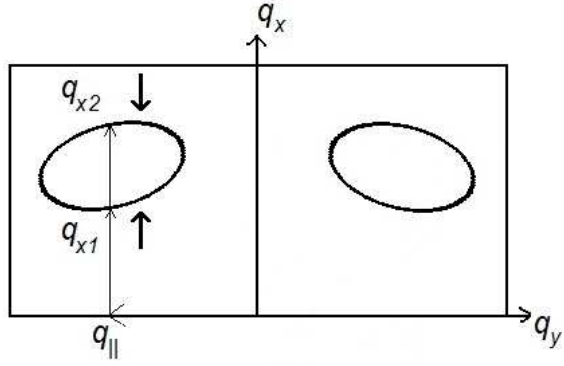}
\caption{ It was shown that to each component $\mathbf{q}_{||}$, lying inside the projection of the isofrequency surface onto the plane $(q_y, q_z)$, correspond two values of $q_x$. To these two values of $q_x$ correspond different values of the derivative $\partial \omega/\partial q_x$, and, hence, different directions of propagation of energy (shown by solid arrows).}
\end{figure}

 The maximum frequency of the transverse oscillations for a simple cubic lattice lies inside and not on the boundary of the Brillouin zone \cite{Ans}. At some frequencies, the surface of constant frequency for these oscillation branches is closed (Fig. 4). In this case to every $\mathbf{q}_{||}$, lying inside the projection of the isofrequency surface onto the plane $(q_y, q_z)$, correspond two values of $q_x$. To these two values of $q_x$ correspond different values of the derivative $\partial \omega/\partial q_x$, and, hence, different directions of propagation of energy. 

     The transmitted waves should have the same sign $q_x$ as the incident (and the reflected the opposite) due to the principle of causality. In addition, because of the energy conservation law, the energy flux direction is the same for the incident and transmitted waves. If we assume that the incident wave is one in which the energy propagates towards the interface, then the value of $q_x$ in such a wave can be both positive and negative. Correspondingly, one should take the sign of $q_x$ in the transmitted and reflected waves. The uncertainty is thus eliminated. 

 \begin{figure}
\centering
\includegraphics[width=0.9\textwidth]{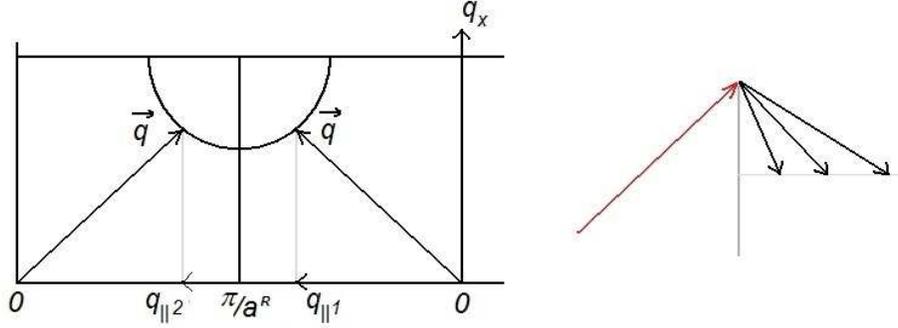}
\caption{ Schematically displayed is the surface of the constant frequency in the Brillouin zone. It was shown that the incident phonon with the parallel-to-interface wave vector component $q_{||}$ lying outside the Brillouin zone of the right crystal, is refracted in the opposite direction. }
\end{figure}

     An interesting effect occurs when the parallel-to-the interface component of the wave vector, $q_{||}$, of the incident phonon is beyond the Brillouin zone of the right crystal (Fig.5). In this case the phonon is refracted and undergoes the Bragg scattering simultaneously. Then it appears that the refracted phonons have the opposite direction of propagation, in the plane of the interface, relative to the incident phonons. Thus, the inverse refraction takes place. 

     It is interesting to note, that the energy flux in the direction perpendicular to the interface should be conserved for each solution of the system of Eqs.(60), and for the solutions of the more exact system of Eqs.(58), (59). However, for each solution, the energy flux in the plane of the interface can be different on the opposite sides from the interface. It means that at the interface not only the temperature may undergo a jump but the thermal flux parallel to the interface plane as well.

\section{Conclusion}

In this paper we have studied the model of the interface between two crystals taking into consideration the mismatch of crystal lattices. The basic assumption of the model is that displacements of atoms near the interface are not random but determined by interaction with atoms of the other crystal. We have shown that in the harmonic approximation such the displacements have no effect on phonon transmission across the interface. The equation was set up, which determines the amplitudes of the transmitted and reflected waves of lattice oscillations. The exact solution has been derived for the quasi-one-dimensional case.

It was shown that the mismatch of lattices leads to the scattering of phonons at the crystal interface. In other words, the scattering appears even in the case of the ideal interface in the absence of defects and roughness. On the other hand, such scattering occurs not uniformly in all directions but has a certain structure.
 
It has been established that the finiteness of the lattice parameter results, at certain angles of incidence, in the backward phonon refraction at the interface. A new family of lattice oscillations has been described, which includes only the atoms lying near the crystal interface. It was predicted that at the plane parallel to the interface, the heat flows suffer a discontinuity at the interface. 

The main result of our study is that we have shown that the oscillations of atoms in the plane of the interface interact weakly with the oscillations of atoms on the opposite side of the interface, except of specific resonance cases. In the case of normal incidence of phonons at the interface, the transmission coefficient of the transversely polarized phonons is much less as compared to that of the longitudinally polarized phonons. For an arbitrary angle of incidence, the transmission coefficients of phonons of any polarization are less than that calculated with the elasticity theory, even in the case of low frequencies. Allowance for this factor leads to that the calculated value of Kapitza resistance is approximately three times greater. Calculation of the interface thermal resistance, performed by means of the method proposed in the work \cite{Me2}, gives significantly lower values as compared to the experiment. Accounting for the smallness of the transmission coefficient of transversely polarized phonons explains this discrepancy.
 
The author thanks E. D. Eidelman and G. V. Budkin for helpful critical comments, and A. Ya. Vul’ for attention to this study. The work was supported by Dynasty Foundation.

\end{document}